\documentclass[aps,prb,reprint,superscriptaddress,longbibliography]{revtex4-1}

\usepackage{todonotes}
\usepackage{amsmath}
\DeclareMathOperator{\sgn}{sgn}
\DeclareMathOperator{\Tr}{Tr}
\DeclareMathOperator{\diag}{diag}

\graphicspath{{../figures/}}

\begin{document}

\date{\today}

\begin{abstract}
We use cellular dynamical mean-field theory with extended unit cells to study the ground state of the two-dimensional repulsive Hubbard model at finite doping. We calculate the energy of states where $d$-wave superconductivity coexists with spatially non-uniform magnetic and charge order and find that they are energetically favoured in a large doping region as compared to the uniform solution. We also study the spatial form of the density and the superconducting and magnetic order parameters at different doping values.

\end{abstract}

\title{A dynamical mean-field theory study of stripe order and $d$-wave superconductivity in the two-dimensional Hubbard model}

\author{Tuomas I. Vanhala}
\affiliation{COMP Centre of Excellence, Department of Applied Physics, Aalto University, Helsinki, Finland}
\author{P\"aivi T\"orm\"a}
\affiliation{COMP Centre of Excellence, Department of Applied Physics, Aalto University, Helsinki, Finland}

\maketitle

\section{Introduction}

The two-dimensional square lattice Hubbard model has been a paradigm model of strongly correlated electron systems for decades. One of the driving forces behind the overwhelming interest to this model has been the idea that it might be the simplest model exhibiting the essential physical principles behind the high-$T_c$ superconductivity of the cuprate materials \cite{lee_doping_2006}. The cuprates are quasi-two-dimensional materials which experimentally are found to exhibit $d$-wave superconductivity and non-uniform magnetic order \cite{vojta_lattice_2009}, both of which have been proposed to exist in the phase diagram of the repulsive Hubbard model for some parameters. Experimental methods for realizing the Hubbard Hamiltonian have also been developed in the ultracold gas community \cite{bloch_many-body_2008,giorgini_theory_2008,torma_quantum_2014}. Especially the recent development of the fermionic quantum gas microscopes provides an interesting platform for observing magnetic order
\cite{mazurenko_cold-atom_2017,drewes_antiferromagnetic_2017,hilker_revealing_2017,parsons_site-resolved_2016,cheuk_observation_2016,cheuk_observation_2016-1,boll_spin-_2016,omran_microscopic_2015,edge_imaging_2015,haller_single-atom_2015},
although further advances are needed to reach the possible superconducting phases.

The idea that the two dimensional repulsive Hubbard model could exhibit superconductivity had been investigated in the context of heavy-fermion superconductors\cite{hirsch_attractive_1985} slightly before the discovery of the cuprate materials, and the number of papers really proliferated after the famous experiment in 1986 \cite{bednorz_possible_1986}. Early exact results from the determinant quantum Monte-Carlo (DQMC) method indicated enhanced $s$-wave\cite{hirsch_attractive_1985} and $d$-wave \cite{hirsch_pairing_1988,white_attractive_1989} pairing susceptibilities.\cite{dagotto_correlated_1994} Unfortunately, at the temperatures and system sizes reachable by DQMC it was not possible to directly demonstrate the existence of a superconducting phase transition, a situation which remains true in the current state of the art \cite{ying_determinant_2014,khatami_finite-temperature_2015}. However, approximative methods, such as the fluctuation-exchange approximation \cite{bickers_conserving_1989}, predicted a $d$-wave superconducting state with a critical temperature of the order of $0.01$ hopping units, more than an order of magnitude lower than what is reached in typical DQMC calculations \cite{khatami_finite-temperature_2015} for the relevant doping regions.

Parallel to the above developments was the invention of the resonating valence bond theory \cite{anderson_resonating_1987,gros_superconductivity_1988,paramekanti_high-$t_c$_2004,anderson_physics_2004}, which, in its simplest version, expresses the ground state using a mean-field-like variational ansatz with the appropriate pairing symmetry projected in the subspace with no double occupancy. Being variational in nature, this theory needs support from other methods, and it is currently not clear how well it can describe the Hubbard model on a quantitative level. Thus obtaining numerical solutions for the Hubbard model is also important because it allows validation of the various approximate theories proposed to explain high-$T_c$ superconductivity. A similar role could be played by the ultracold gas experiments, where simple model systems can be realized and studied accurately \cite{bloch_quantum_2012}.

Soon after the discovery of the superconducting cuprate compounds, experiments indicated possible incommensurate magnetic order in the hole-doped region \cite{birgeneau_static_1989}.
Motivated by this, it was found that \cite{zaanen_charged_1989,poilblanc_charged_1989,machida_magnetism_1989,kato_soliton_1990}, within mean-field theory, variants of the $2$D Hubbard model could also support a so-called \emph{stripe} state, where the magnitude of the magnetization is spatially modulated.
It has not always been recognized that such modulated magnetization is closely related to the famous FFLO superconducting state of the attractive model, predicted within mean-field theory already earlier \cite{kinnunen_fulde-ferrel-larkin-ovchinnikov_2017}. Indeed, on a bipartite lattice, the FFLO state can be mapped into a stripe state via a simple particle-hole transformation \cite{ho_quantum_2009,kinnunen_fulde-ferrel-larkin-ovchinnikov_2017}. In finite size DQMC calculations one can observe short range stripe order and enhanced susceptibilities \cite{huang_stripe_2017,wolak_pairing_2012}, but, largely due to the fermionic sign problem, a finite size scaling analysis clearly showing a phase transition to a striped state is still out of reach.

The indications that the electronic correlations play an important role in the cuprates and the advances in the ultracold gas experiments have motivated the development of a variety of new numerical methods for lattice models. Results indicating the existence of the FFLO state in the 2D Hubbard model at a finite temperature were recently obtained using an unbiased diagrammatic Monte-Carlo approach \cite{gukelberger_fulde-ferrell-larkin-ovchinnikov_2016}, giving a maximum critical temperature of the order of $0.1$ hopping units. Various ground state methods have also been developed. These include the density matrix embedding theory (DMET) \cite{knizia_density_2012}, constrained path auxiliary field Monte-Carlo (AFQMC) \cite{zhang_constrained_1995}, the tensor network wave function ansatz based iPEPS \cite{jordan_classical_2008}, and advances in the density matrix renormalization group method (DMRG) to handle wider quasi-2D systems \cite{stoudenmire_studying_2012}. A study comparing these four methods consistently found striped states at $\frac{1}{8}$ doping \cite{zheng_stripe_2016}. The DMET has also been used to map the density-interaction phase diagram of the Hubbard model showing (co)existence of stripe states and $d$-wave superconductivity \cite{zheng_ground-state_2016}, and AFQMC has been used to study long wavelength stripes at low doping \cite{chang_spin_2010}. The existence of the stripe state is also being debated for the closely related $t$-$J$-model, where stripes have been studied using iPEPS \cite{corboz_competing_2014}, variational monte carlo \cite{himeda_stripe_2002}, and fixed-node Monte-Carlo (FN) methods \cite{hu_absence_2012}.

The dynamical mean-field theory (DMFT) with its cluster extensions is another relatively recent, although already well established, numerical method \cite{georges_dynamical_1996,kotliar_electronic_2006,maier_quantum_2005}. It has been used to gain insight to the Mott transition both in the Hubbard model \cite{georges_dynamical_1996} and in real materials \cite{kotliar_electronic_2006}. Since the initial realization \cite{lichtenstein_antiferromagnetism_2000} that the plaquette DMFT (meaning cellular DMFT \cite{maier_quantum_2005} with a $2 \times 2$ cluster) can describe $d$-wave superconductivity, this approximation has been employed to study the competition and coexistence of superconductivity and magnetic order in the ground state \cite{capone_competition_2006,kancharla_anomalous_2008}, the effect of nearest neighbour repulsion \cite{senechal_resilience_2013,jiang_d-wave_2017}, and finite temperature energetics and critical temperatures of the superconductivity \cite{kitatani_flex+dmft_2015,fratino_organizing_2016}. This approximation in its basic form cannot produce a striped state, since the $2 \times 2$ unit cell does not allow modulation of the magnetization across the lattice.

The plaquette DMFT can be regarded as a lowest order mean-field approximation for the $d$-wave superconductor, as it does not take into account correlations beyond neighbouring sites. However, large scale cluster DMFT calculations \cite{maier_systematic_2005,gull_superconductivity_2013,staar_two-particle_2014} have provided evidence that the superconductivity exhibited by the plaquette DMFT is not just an artefact of the approximation, but indeed exists also for larger clusters.
The stripe and FFLO states have also been investigated within real-space DMFT calculations where correlations do not extend beyond a single site
\cite{peters_spin_2014,heikkinen_finite-temperature_2013,kim_fulde-ferrell-larkin-ovchinnikov_2012,raczkowski_melting_2010,fleck_one-dimensional_2000}, but inhomogeneous solutions are allowed. Such approximations can not produce $d$-wave superconductivity, where the pairing is induced by correlations between different sites, which also holds for simpler Hartree-Fock mean-field theory. A cluster DMFT study showing resilience of the FFLO state against long range quantum fluctuations in a quasi-1D setting also exists \cite{heikkinen_nonlocal_2014}.

In light of the numerical evidence and the physical understanding from mean-field methods, the existence of the stripe state in the repulsive Hubbard model at least for some parameter region seems highly plausible. However, its interplay with the $d$-wave superconductivity has not yet been thoroughly explored. At the fixed doping $\frac{1}{8}$, superconducting states were found within DMET, iPEPS and DMRG, but the states seemed to be metastable \cite{zheng_stripe_2016}. The topic has also been touched in other DMET \cite{zheng_ground-state_2016} and functional renormalization group calculations \cite{yamase_coexistence_2016}, as well as in variational calculations of finite width chains \cite{leprevost_intertwined_2015,zheng_ground-state_2016}, but has not been taken into consideration in the DMFT-based Green's function methods. In addition, an analysis of the energetics for the different stripe wavelengths for a wide region of doping values has been missing. Several works have also investigated superconductivity coexisting with inhomogeneities induced by various external potentials \cite{okamoto_microscopic_2010,goren_enhancement_2011,mondaini_determinant_2012,maier_dynamic_2010}, but it is not clear if such studies are relevant for the stripe states with spontaneous translational symmetry breaking.

In this article we present plaquette DMFT calculations with extended unit cells and study the interplay of $d$-wave superconductivity and striped magnetic order. Thus, on the one hand, we extend the single-site real-space DMFT calculations \cite{peters_spin_2014,heikkinen_finite-temperature_2013,raczkowski_melting_2010,fleck_one-dimensional_2000} to include $d$-wave superconductivity and, on the other, we extend the uniform plaquette DMFT approximation \cite{lichtenstein_antiferromagnetism_2000,capone_competition_2006,kancharla_anomalous_2008,senechal_resilience_2013,kitatani_flex+dmft_2015,fratino_organizing_2016} to include the striped order. We find a wide region of coexistence between striped order and spatially modulated $d$-wave superconductivity, and analyze the behaviour of the superconducting order parameter in the striped states. Compared to other recent ground state studies, we can reach longer stripe wavelengths \cite{zheng_ground-state_2016} and study a wider doping region \cite{zheng_stripe_2016}. This allows us to more carefully study the stripe energetics and to determine the ground state stripe wavelengths.

\section{The Model and the method}

We study the square lattice Hubbard model, whose grand-canonical Hamiltonian can be written in the particle-hole symmetric form
\begin{equation}
\begin{split}
H_{gc} &= t\sum_{ \left\langle ij \right\rangle \sigma} c_{i\sigma}^\dagger c_{j\sigma} + \mu \sum_i (n_{i\uparrow} + n_{i\downarrow}) \\
         & + U \sum_i \left(n_{i\uparrow}-\frac{1}{2} \right) \left(n_{i\downarrow}-\frac{1}{2} \right) \\
&= H + \left(\mu-\frac{U}{2} \right) \sum_i (n_{i\uparrow}+n_{i\downarrow}) - \sum_i \frac{U}{4},
\end{split}
\label{hamiltonian}
\end{equation}
where $\mu$ is a chemical potential, $U$ the interaction strength, $n_{i\sigma}$ the density operator of spin $\sigma$ at site $i$, and $\left\langle ij \right\rangle$ denotes nearest neighbour hoppings. Here we also define $H$ as the Hamiltonian without the chemical potential and constant terms. We measure energies in terms of the hopping parameter $t$, and concentrate here on the intermediate interaction strength $U=6t$.

Our goal is to study the interplay of the stripe order and $d$-wave superconductivity in the ground state of this model. For this purpose we use cellular dynamical mean-field theory \cite{maier_quantum_2005} (CDMFT) with extended unit cells. Dynamical mean-field theory treats short range quantum correlations using a many-body impurity model, which is solved exactly, and includes the lattice physics using a self-consistency scheme. The approximation is defined directly in the limit of an infinite lattice, by approximating the self-energy of the lattice system using the self-energy of the impurity model. In the CDMFT formulation the lattice is divided into clusters consisting of a number of adjacent sites, and the approximated self-energy is local within each cluster. To study the superconducting solutions, we use the Nambu-formalism, where the Green's function of the problem is written in the Nambu-spinor notation
\begin{equation}
G_{ij}(\tau) = \left\langle \mathcal{T} \psi_i(\tau) \psi_j(0)^\dagger \right\rangle, 
\psi_i(\tau)=
\begin{bmatrix}
c_{i\uparrow}(\tau) \\
c_{j\downarrow}^\dagger(\tau)
\end{bmatrix}
\end{equation}
As the self-energy of a $d$-wave superconductor is non-local, the single-site DMFT is not enough. Here we use the smallest possible cluster size to produce the $d$-wave state, which is the $2 \times 2$ cluster known as the plaquette.

\begin{figure*}
\includegraphics[width=1.9\columnwidth]{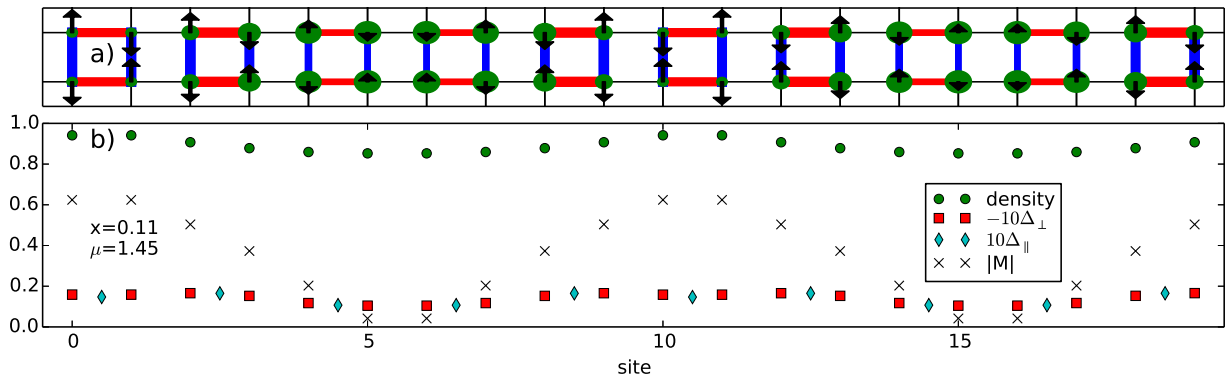}
\caption{a) Schematic representation of the unit cell for $\lambda=10$ at doping $x \approx 0.11$. In this case the unit cell is of length $20$ and includes $N_c=10$ clusters. In the plaquette DMFT approximation the self-energy is local within each cluster. The arrows show the magnetization for each site and the thickness of the red and blue lines indicates the magnitude of the order parameters $\Delta_\perp$ (vertical) and $\Delta_\parallel$ (horizontal) within each cluster. Red color indicates a positive order parameter, and blue a negative one. The radius of the dots is scaled according to the local doping (i.e. hole density). b) Density, the superconducting order parameters and the absolute value of the magnetization for the unit cell in panel a. In the stripe state the staggered magnetization is spatially modulated and in this example case changes sign between the third and the fourth cluster and again between the eigth and the ninth cluster. The sign changes correspond to magnetic domain walls between neighbouring stripes. \label{LatticeDiagramFigure}}
\end{figure*}

In previous CDMFT studies \cite{capone_competition_2006,fratino_organizing_2016,lichtenstein_antiferromagnetism_2000} it has been assumed that the self-energy of each cluster is equal, producing a uniform $d$-wave superconducting state with the same density and same staggered magnetization throughout the lattice. In our case we relax this assumption by considering larger unit cells of size $2 \times 2N_c$, where $N_c$ is the number of the $2 \times 2$ clusters within the unit cell. This approach is reminiscent of the so-called real space DMFT \cite{snoek_antiferromagnetic_2008}, where every site of a finite lattice produces an inequivalent impurity problem. In our case every $2 \times 2$ cluster in the unit cell corresponds to an independent (cluster) impurity problem that has to be solved at each iteration of the self-consistency loop. This allows the strength of the $d$-wave pairing and the magnetization to vary within the unit cell, as depicted in figure \ref{LatticeDiagramFigure}a. Because the unit cell is always of even length, we only study stripe states of even periodicity. For compatibility with existing numerical literature \cite{zheng_stripe_2016}, we define the stripe wavelength $\lambda$ as half the wavelength of the staggered magnetization, so that $\lambda=N_c$.

To obtain the stripe ordered state we take a frequency independent initial guess for the self-energy chosen so that it produces a staggered magnetization of approximately sinusoidal form with wavelength $2N_c$. In addition we break the particle conservation symmetry by adding a small symmetry breaking term $\sum_{ij} (\delta_{ij} c_{i\uparrow} c_{j\downarrow} + h.c.)$ to the Hamiltonian. Here $\delta_{ij}$ is chosen to be of magnitude $0.001$ and to have the $d$-wave symmetry within each cluster (see Fig. \ref{LatticeDiagramFigure}). This is needed, as otherwise the iteration is stuck in the normal state solutions where the anomalous component of the Green's function $\left\langle c_{i\uparrow} c_{j\downarrow} \right\rangle$ is zero. After perfoming the self-consistency loop for some number of iterations, we set $\delta$ to zero and run the loop until convergence. We note that the chosen initial guess effectively fixes the wavelength of the obtained solution. For example, a unit cell with $N_c=16$ is able to accommodate stripes with both $\lambda=8$ and $\lambda=16$, but the self-consistency iteration is usually stuck with whatever $\lambda$ was initially chosen.

To find out the stripe wavelength giving the ground state of the system at a given doping, we calculate the total energy of the system from the Matsubara self-energy $\Sigma$ and Green's function $G$ of the converged results using the formula \cite{kotliar_electronic_2006}
\begin{equation}
\left\langle H_{gc} \right\rangle=\sum_{n} \Tr \left( H_0 G(i \omega_n) + \frac{1}{2} \Sigma(i\omega_n) G(i \omega_n) \right),
\end{equation}
where $\Sigma$ and $G$ are matrices in spinor- and site-indices, and $H_0$ is the single-particle grand canonical Hamiltonian matrix. In practice the trace is evaluated in momentum space. Furthermore, we use a fitting procedure to find the lowest moments of the $\frac{1}{\omega_n}$ expansion of the trace, and calculate the contribution from the high frequency tail analytically, as is commonly done for the Fourier transform of $G$ itself. Afterwards we substract the chemical potential and constant terms to obtain the energy $E=\left\langle H \right\rangle$.

To solve the impurity problem, we use the exact diagonalization (ED) method \cite{georges_dynamical_1996,capone_cluster-dynamical_2004} with 16 bath orbitals (or 8 bath sites). We discretize the Green's functions using a fictitious inverse temperature $\beta=50$. The single-particle Hamiltonian of the cluster problem can be expressed in the form
\begin{equation}
H=H_c + \sum_{ij} (\psi_{i}^\dagger h_{ij} \phi_{j} + \phi_{j}^\dagger h_{ij} \psi_{i}) + \sum_j \phi_{j}^\dagger \diag(\epsilon_{j\uparrow},\epsilon_{j\downarrow} ) \phi_j,
\label{ImpurityProblemNoninteractingHamiltonian}
\end{equation}
where the Nambu spinors $\psi$ represent the cluster sites, and spinors $\phi$ represent the bath sites. We choose $H_c$ to be the cluster hopping Hamiltonian, which is the lattice Hamiltonian restricted to within the cluster. The bath parameters $h_{ij}$ and $\epsilon_j$ are determined using an unrestricted fitting procedure that finds a minimum of the cost function
\begin{equation}
C(h,\epsilon)=\sum_{ijn} \frac{1}{|\omega_n|} \left| \mathcal{G}_{ij}(i\omega_n) - G^{c0}_{ij}(i\omega_n) \right|^2,
\end{equation}
where $|\cdot|$ is the Frobenius norm, $\mathcal{G}$ is the CDMFT bath Green's function and $G^{c0}$ is the non-interacting cluster Green's function of the Hamiltonian in equation \ref{ImpurityProblemNoninteractingHamiltonian}, both in the Nambu form. The interacting Green's function of the impurity problem is solved using a band Lanczos procedure \cite{senechal_introduction_2008,freund_band_2000}. Denoting the numbers of up- and down-particles in the impurity problem by $N_{\uparrow}$ and $N_{\downarrow}$, we note that the conserved particle number, $\sum_i \psi_i^\dagger \psi_i + \sum_j \phi_j^\dagger \phi_j = 12+N_\uparrow - N_\downarrow$ is always $12$, as the total polarization of each cluster is zero. (This quantity is the number of $\uparrow$ particles plus the number of $\downarrow$ holes on the $4+8$ sites of the impurity problem.) This saves a scan over the particle number sectors when finding the ground state. We also note that the ED method cannot practically handle larger clusters than the plaquette while still retaining a reasonable number of bath orbitals, and thus we cannot study the effect of the cluster size within this method.

To probe the superconductivity we define the order parameter $\Delta_b$ associated with the lattice bonds $b=(i,j)$ as
\begin{equation}
\Delta_b=\left \langle c_{\uparrow i} c_{\downarrow j} \right\rangle + \left \langle c_{\downarrow i} c_{\uparrow j} \right\rangle
\end{equation}
Here we calculate this order parameter for the different lattice bonds within the CDMFT clusters. For each cluster we have two equivalent bonds in the long direction of the unit cell, which give the same order parameter labelled by $\Delta_\parallel$ (see figure \ref{LatticeDiagramFigure}). In addition we have two inequivalent bonds in the short direction, which we label as $\Delta_\perp$. For the $d$-wave superconductor the order parameters $\Delta_\parallel$ and $\Delta_\perp$ have opposite signs, and the local order parameter $\left \langle c_{\uparrow i} c_{\downarrow i} \right \rangle = 0$. Furthermore, we measure the local staggered magnetization which is defined as
\begin{equation}
M_i=\sgn(i)(n_{i \uparrow} - n_{i \downarrow}),
\end{equation}
where $\sgn(i)$ is a sign that is opposite for neighbouring sites. We discuss the results in terms of the doping $x$ defined as
\begin{equation}
x=1 - \rho,
\end{equation}
where $\rho \in [0,2]$ is the total number of particles per site.

\section{Results}

\begin{figure}
\includegraphics[width=0.95\columnwidth]{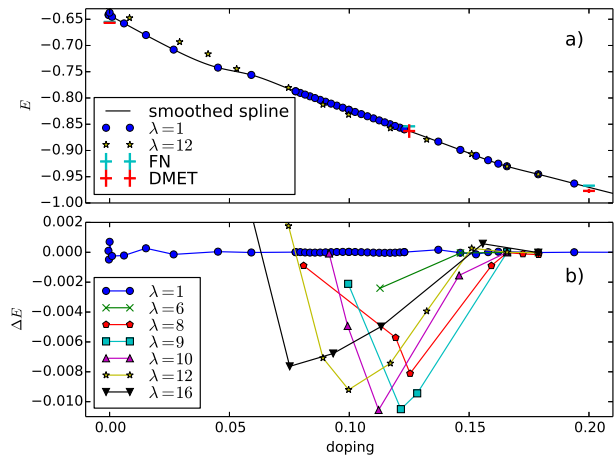}
\caption{a) Energy of the uniform state and the stripe state of wavelength $\lambda=12$ as a function of the doping. The DMET and FN results and their error estimates (indicated by the vertical bars) are from tables II, IV and V of LeBlanc et al \cite{simons_collaboration_on_the_many-electron_problem_solutions_2015}, and represent extrapolations to infinite (embedded) system size. b) Energy for different stripe wavelenghts relative to the smoothed spline interpolant fitted to the uniform case energy. \label{EnergyFigure}}
\end{figure}

\begin{figure}
\includegraphics[width=0.95\columnwidth]{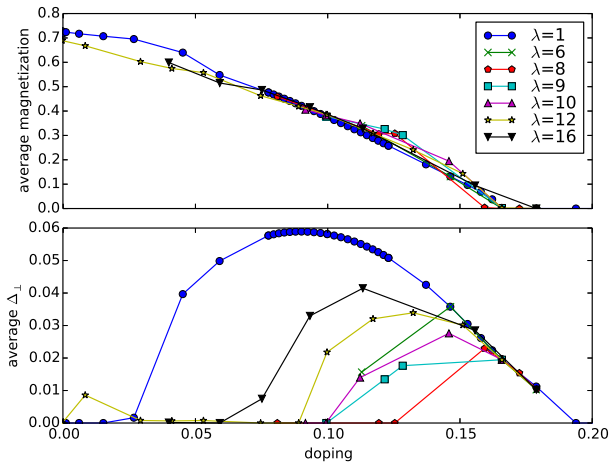}
\caption{The average staggered magnetization (upper panel) and the average $d$-wave order parameter $\Delta_{\perp}$ (lower panel) as a function of the doping. \label{OrderParameterFigure}}
\end{figure}

Here we will first discuss the energetics of the different stripe wavelengths, and then move to a discussion of the superconducting order parameters. Figure \ref{EnergyFigure}a shows the energy for the different stripe wavelengths as a function of the doping. The energy is defined as the expectation value of the Hamiltonian $H$ in equation \ref{hamiltonian} which does not include the chemical potential terms. For comparison we have included some results from density matrix embedding theory (DMET) and fixed-node diffusion Monte Carlo (FN) from LeBlanc et al \cite{simons_collaboration_on_the_many-electron_problem_solutions_2015}. These results have been extrapolated to the thermodynamic limit in embedded cluster size (DMET) or system size (FN), while our calculations are performed for the $2x2$ cluster. The DMET error estimates include all sources of error, while the (very small) FN error bars do not include errors from the fixed node approximation \cite{simons_collaboration_on_the_many-electron_problem_solutions_2015}. In absolute terms the energies are in reasonably good agreement with the other methods, which gives confidence that the energy calculation procedure is technically correctly implemented and reliable. We note, however, that the differences between the different methods are generally larger than the differences in the energies of the different wavelength striped solutions discussed below.

In figure \ref{EnergyFigure}b we plot the energy relative to the energy of the uniform state as a function of doping. 
It can be observed that the wavelength that gives the lowest energy depends on the doping and increases towards half-filling, which is expected from mean-field theory \cite{kinnunen_fulde-ferrel-larkin-ovchinnikov_2017}. Zheng et al. \cite{zheng_stripe_2016} studied the same model using different numerical methods at doping $0.125$. There it was found that the lowest energy stripe state is given by a wavelength of $\lambda \approx 8$ for $U=6$ and $U=8$. Our results indicate the stripe of wavelength $\lambda=9$ to have the lowest energy for the corresponding doping, although $\lambda=8$ is also close in energy. It should be noted that the optimal stripe wavelength at a given doping is not necessarily commensurate with the lattice, so that the results have to be treated as best integer approximations to the correct stripe wavelength.

In the single-site inhomogeneous (or real-space) DMFT study \cite{peters_spin_2014}, which does not include superconductivity, a relation $\lambda_{gs}=1/x$ was demonstrated for the ground state wavelength $\lambda_{gs}$ and doping $x$. Our results again indicate perhaps slightly longer ground state wavelengths, such as $\lambda_{gs}=9$ for $x=1/8$ and $\lambda_{gs}=10$ for $x=1/9$. This might be because the longer wavelength is better compatible with superconducting order, as discussed below, thus allowing the system to lower its energy. However, the differences are quite small, and the relation $\lambda_{gs}=1/x$ is a good approximation also with the superconductivity included.

Figure \ref{EnergyFigure} shows that the largest difference in energy between the uniform state and the striped state is obtained somewhere close to doping $0.11$ with the best stripe wavelength being close to $10$. This is an interesting observation, as the point with the largest energy difference could be the best point to start searching for definite signatures of the stripe order in e.g. ultra cold gas experiments. We also note that the state-of-the-art experiments can already achieve system sizes with linear dimension of the order of $10$ lattice sites in a uniform potential \cite{mazurenko_cold-atom_2017}.

\begin{figure}
\includegraphics[width=0.8\columnwidth]{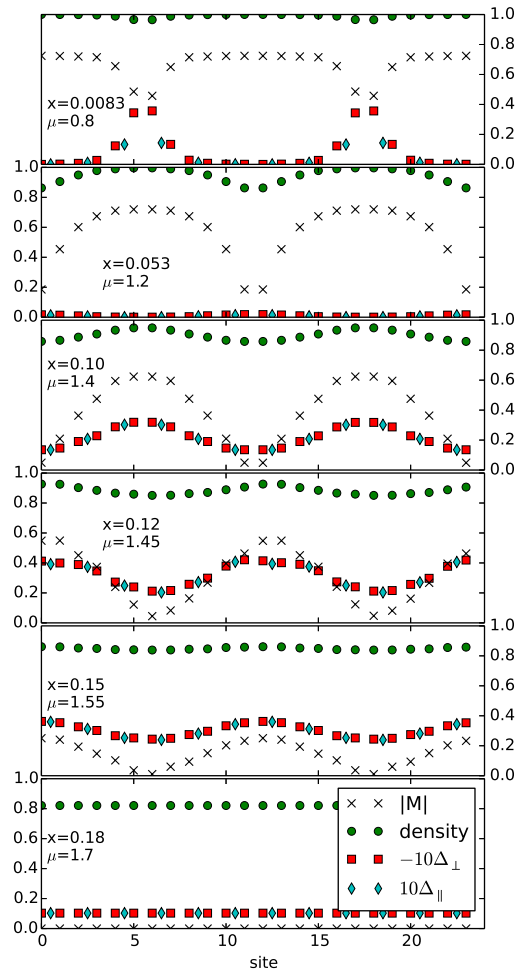}
\caption{The staggered magnetization, density and $d$-wave order parameters for $\lambda=12$ and different dopings $x$. Doping $x=0.10$ gives the lowest energy relative to the uniform state, while the other cases represent metastable states (cf. figure \ref{EnergyFigure}). \label{SpatialVariationFigure}}
\end{figure}

Next we will discuss the behaviour of the superconducting order parameter as a function of the doping and stripe wavelength. In figure \ref{OrderParameterFigure} we plot the average staggered magnetization and $d$-wave order parameter as a function of the doping. This reveals a pattern where the extent of the superfluid region grows when the stripe wavelength is increased, but it still always remains smaller than for the uniform ansatz. This does not necessarily mean that the region of superconductivity in the ground state would be smaller when the non-uniform states are taken into account, because the ground state wavelength grows when approaching half-filling. Indeed, from our data it seems that the state with the lowest energy at a given doping always carries superconductivity, except in regions where even the uniform state has vanishing superconducting order parameter.

In earlier work \cite{zheng_stripe_2016} for the doping $0.125$ it was found that there is no coexistence of superconductivity and $d$-wave order in the ground state ($\lambda=8$) stripe at $U=8$. However, using iPEPS and DMRG, superconductivity was found for different stripe lengths, such as $\lambda=5$ and $\lambda=7$, which were also very close in energy. We note that also in our results the superconducting order parameter for the $\lambda=8$ stripe is small or zero in the vicinity of doping $0.125$, but we find the $\lambda=9$ stripe to have essentially the same or slightly lower energy and to carry superconductivity. We note that it is difficult to obtain converged solutions for the $\lambda=8$ stripe in the vicinity of the magnetic phase transition close to $x=0.15$, and the result depends strongly on tiny changes in the chemical potential. From partially converged results (not shown in the plots) it still seems that there is superconductivity for $\lambda=8$ around this doping value.

Zheng et al \cite{zheng_ground-state_2016} have also performed DMET calculations to study the $U$-doping phase diagram. At $U=6$ they find antiferromagnetic order at half-filling, and superconducting order partially coexisting with inhomogeneous magnetic order roughly from doping $0.1$ to $0.3$. 
Qualitatively, the most notable difference between their results and ours is that they find an intermediate metallic point at about doping $0.1$, between the antiferromagnetic and superconducting regions, where both magnetic order and superconductivity vanish, while we do not see indications of this. However, we also note that for uniform plaquette DMFT superconductivity and magnetic order do not coexist for large interactions $U \gtrsim 8$, instead having a first order transition between them \cite{capone_competition_2006}, which is perhaps more compatible with the DMET result at $U=6$. It thus seems that a more useful qualitative comparison between the various methods should be performed with the full $U$-doping phase diagram at hand, which is an important goal for future work. We also note that the DMET calculation employed unit cells allowing a maximum wavelength of $\lambda=4$.

We can also discuss the development of the density and the order parameters as a function of the doping for the fixed wavelength $\lambda=12$, see figure \ref{SpatialVariationFigure}. The case $x=0.10$ is the point where this stripe wavelength gives the lowest energy state relative to the other wavelengths. When the doping is increased from $0.10$, the magnetization decreases and eventually goes to zero, at which point the state joins the uniform $d$-wave superconducting state, which in turn transitions into the normal state at about doping $0.19$. When going to the opposite direction, the density first increases along the whole unit cell, until the high-density regions reach half-filling. At this point the system is in a metastable state with regions of uniform magnetization and ``domain walls'' where the direction of the staggered magnetization changes. For suitable parameters close to half-filling it is possible to obtain metastable solutions where superconductivity is present at these domain walls (shown in the uppermost panel of figure \ref{SpatialVariationFigure}).

Returning to the ground state case $x=0.10$, we can see that the largest value of the magnetization corresponds to the maximum of the density, which is expected as the magnetization of the uniform solution is largest close to half-filling. It is notable that the maximum of the $d$-wave order parameter also occurs at the same sites as the maximal magnetization. This is in contrast to proposals that the cuprate superconductors could have a stripe state where the superconductivity occurs at regions of lower electron density and magnetization. The situation is reversed in the case of $\lambda=6$ at doping $0.11$, where the maximal $d$-wave order parameter occurs at the minimum of the magnetization. While this state is not the ground state in the present case, it is still conceivable that the ground state exhibits this behaviour in some part of the $U$-doping phase diagram. This has been seen e.g. in studies of the $t$-$J$ model, using a variational tensor network ansatz for the ground state \cite{corboz_competing_2014}.

\begin{figure}
\includegraphics[width=0.8\columnwidth]{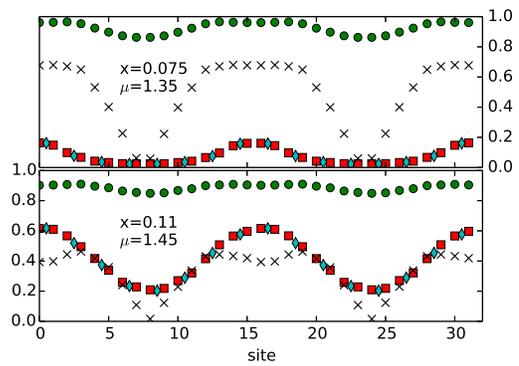}
\caption{Order parameters for $\lambda=16$. The symbols are as in figure \ref{SpatialVariationFigure}. \label{SpatialVariationFigure16}}
\end{figure}

For the largest unit cell with $\lambda=16$ the form of the order parameters is noticeably different from the case $\lambda=12$. For the case $x=0.075$ (upper panel in figure \ref{SpatialVariationFigure16}) we obtain a solution with regions of nearly uniform magnetization and high superconducting order parameters and transition regions between them. The shape of the magnetic order parameter resembles the ``strong FFLO'' order found in a mean-field study of the three-dimensional attractive model \cite{loh_detecting_2010} and in the single-site inhomogeneous DMFT study \cite{peters_spin_2014} for low doping values. Also in this case we note that the superconducting order parameter has maxima within the strongly magnetized regions. For higher dopings it is difficult to obtain well-converged solutions for this largest unit cell, and the obtained shapes of the magnetic order parameters seem somewhat irregular (see the lower panel of figure \ref{SpatialVariationFigure16}). This seems natural, however, since the ground state wavelength is actually shorter.

\section{Discussion and Conclusions}

In this article we have studied the coexistence of $d$-wave superconductivity and striped magnetic and charge order in the ground state of the two-dimensional Hubbard model. We have employed a plaquette DMFT approximation with expanded unit cells which allows the formation of both inhomogeneous magnetic states and $d$-wave superconductivity. Our study thus extends existing single-site real-space DMFT studies of the stripe state and plaquette DMFT studies of the $d$-wave superconductivity by simultaneously allowing both types of order. The advantage of this real-space treatment over the full cluster and finite-size size methods employed e.g. by Zheng et al \cite{zheng_ground-state_2016,zheng_stripe_2016} is that the computational time for a single iteration of the self-consistency loop scales only linearly with the unit cell size, allowing us to study stripe wavelengths up to $\lambda=16$ for multiple values of the doping. The approximation neglects correlation effects beyond neighbouring sites, and thus it can be regarded as a type of mean-field solution for the $d$-wave superconductivity.

Our results are generally consistent with previous studies. The uniform solution is similar to those in earlier work \cite{capone_competition_2006,kancharla_anomalous_2008}, although we did not study exactly the same parameters. The plaquette-DMFT energy is in reasonable agreement with recent numerical results \cite{simons_collaboration_on_the_many-electron_problem_solutions_2015}, although the differences between methods are generally larger than e.g.~differences in the energies of different wavelength stripe states. Some systematic errors are naturally expected because of the mean-field nature of the method. At half-filling, where high-accuracy reference data is available \cite{simons_collaboration_on_the_many-electron_problem_solutions_2015}, plaquette-DMFT overshoots the ground state energy by about $0.01$ hopping units. Away from half-filling, agreement between the different reference methods is not sufficient to judge the magnitude of the error. The lowest energy stripe wavelengths are close to those in the single-site DMFT study \cite{peters_spin_2014}, and the study \cite{zheng_stripe_2016} comparing multiple methods at doping $x=1/8$, although we tend to get slightly longer wavelengths such as $\lambda=9$ instead of $\lambda=8$ at $x=1/8$.

The main results of our study concern the behaviour of the superfluid order in the striped states. Some doping region where the superconducting and stripe states coexist is found for all studied stripe wavelengths. However, we find that the superconductivity is suppressed relative to the uniform solution: the striped solutions produce both smaller values of the order parameter and a smaller range of dopings where the superconductivity exists at all. When the stripe wavelength is increased, the region of superconductivity extends to lower dopings, as shown in figure \ref{OrderParameterFigure}. It seems reasonable to assume, that very long wavelength solutions would give essentially the same superconducting region as the uniform solution. Whether the actual ground state is superconducting whithin the present approximation must be determined by looking at the stripe wavelength giving the lowest energy. From the numerical data it seems that the superconductivity indeed persists in the ground state despite the suppressing effect of the stripe order.

Studying the spatial structure of the order parameters, we find that for the optimal stripe wavelengths the maximal superconducting order parameter tends to occur at the same position as the maximal magnetization, which is in contrast to studies of the $t$-$J$ model \cite{corboz_competing_2014,tu_genesis_2016}. This is true also for the longest wavelength stripes where superconductivity occurs within regions of strong, essentially uniform magnetization. As the $t$-$J$ model is related to the strong coupling limit of the Hubbard model, it is an interesting qualitative question how this behaviour is affected by the interaction strength.

In this work we have studied the so-called vertical stripes, where the direction of the spatial modulation is along the lattice bonds. This is the type of stripe state that is usually found to have lowest energy \cite{zheng_stripe_2016,corboz_competing_2014,corboz_improved_2016}, although the difference to striped order in the diagonal direction can be very small \cite{peters_spin_2014,corboz_competing_2014,corboz_improved_2016}. Similarly, for the $d$-wave superconductivity we have used an ansatz of the in-phase type, where $\Delta_{\perp}$ has the same sign throughout the whole unit cell. In previous studies of the Hubbard and $t$-$J$ models where $d$-wave order is found, this type is usually found to be most stable \cite{zheng_stripe_2016,zheng_ground-state_2016,corboz_competing_2014}, although the antiphase type where a sign change occurs can also be close in energy \cite{corboz_competing_2014}.

An important topic for future research is the finite temperature behaviour of the striped phases coexisting with the superconductivity. In this respect the dynamical mean-field theory based methods have an advantage over the simpler wave function based ones, which are formulated for ground state calculations. Thus we hope that the present work can function as a starting point for finite temperature studies. However, it must be kept in mind that, especially in two dimensions, long range thermal fluctuations (of the phase of the superconducting order parameter and possibly the wavelength and direction of the stripes) neglected in CDMFT are expected to be important for finite temperature physics. This is especially true when the different stripe wavelengths and even different types of order seem to be very close in energy. Thus, an effective theory of such fluctuations is probably required, even though the parameters of such a theory, such as the superfluid phase stiffness, might be evaluated using (dynamical) mean-field methods.

\acknowledgments

We thank Pramod Kumar, Long Liang and Ari Harju for useful discussions. This work was supported by the Academy of Finland through its  Centers  of  Excellence  Programme  (2012-2017)  and under  Project  Nos. 284621, 303351 and 307419, and by  the  European  Research  Council  (ERC-2013-AdG-340748-CODE). T.I.V. acknowledges the support from the V\"ais\"al\"a foundation.
Computing resources were provided by CSC - the Finnish IT Centre for Science.

\appendix

\section{Order parameter profiles for different doping values}

\begin{figure*}
\includegraphics[width=2\columnwidth]{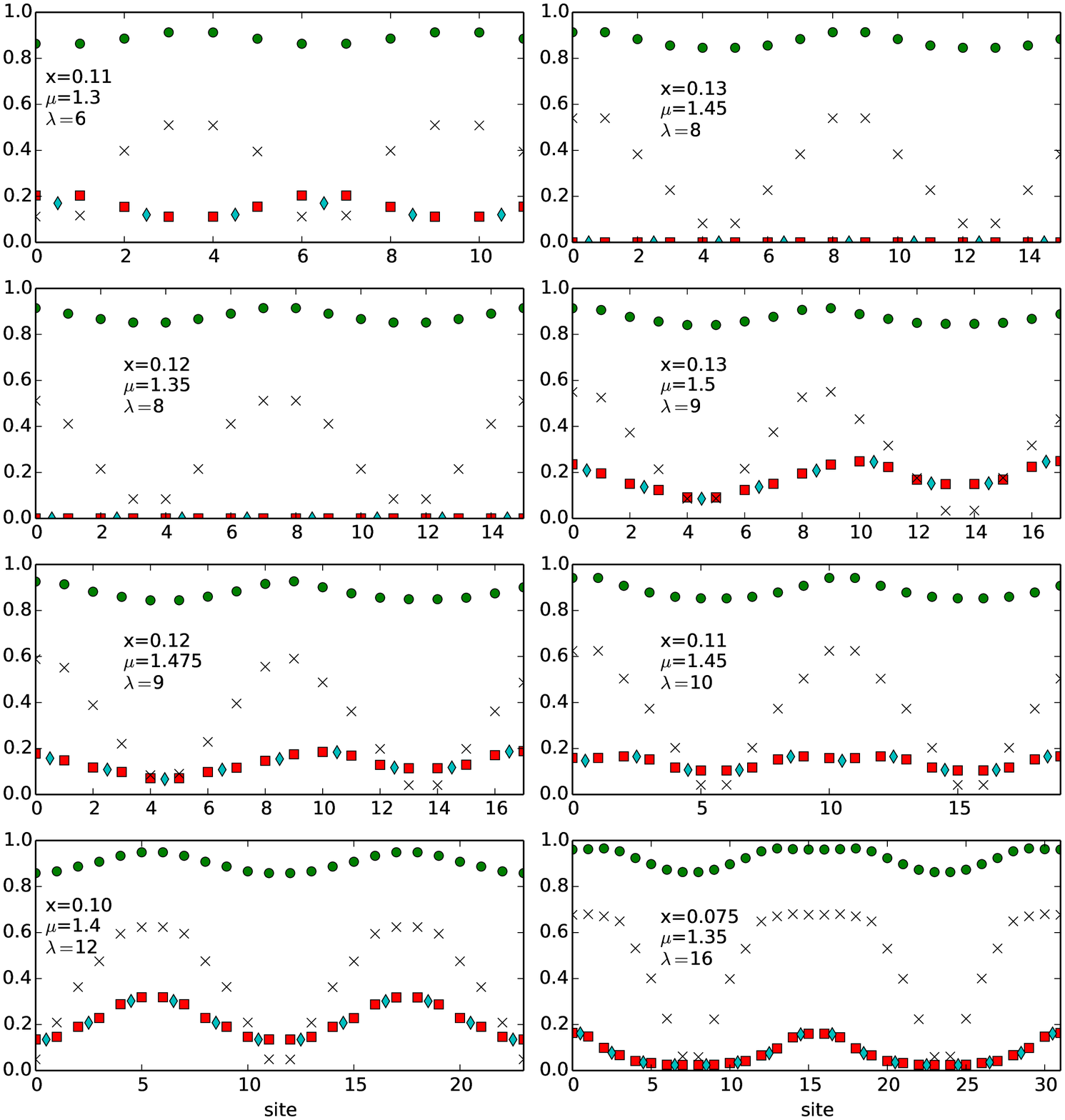}
\caption{Order parameters for different wavelengths. The two panels in the top row and the left panel in the second row show non-ground-state cases. The rest of the plots represent lowest-energy states for dopings decreasing down and to the right. The symbols are as in figure \ref{SpatialVariationFigure}. Note that the wavelength $10$, $12$ and $16$ cases were plotted in other figures as well, but are reproduced here for comparison. \label{GroundStateFigure}}
\end{figure*}

Here we provide some additional comparisons of the spatial form of the order parameters for different wavelengths. Figure \ref{GroundStateFigure} shows the  ground state at different dopings as well as two cases of higher energy states. For $\lambda=6$ we obtain a solution where the highest superconducting order parameter appears at the point of lowest magnetization.

\bibliography{DWaveAndStripes.bib}

\end{document}